\documentclass[twocolumn,groupedaddress,showpacs]{revtex4}
\usepackage{graphicx,subfigure,amsmath, amsthm, amssymb,textcomp,color}

\usepackage{natbib}
\begin{document}
\title{Optimisation of fractal spaceframes under gentle compressive load}
\author{Daniel Rayneau-Kirkhope$^{1}$, Yong Mao$^2$, Robert Farr$^{3, 4}$}
\affiliation{
$^1$ Aalto Science Institute, School of Science, Aalto University, 02150 Espoo, FINLAND\\
$^2$ School of Physics and Astronomy, University of Nottingham, Nottingham, NG7 2RD, UK \\
$^3$ Unilever R\&D, Colworth House, Sharnbrook, Bedford, MK44 1LQ, UK\\
$^4$ London Institute for Mathematical Sciences, 35a South Street, Mayfair, London, UK}
\date{26 June 2013}
\begin{abstract}
The principle of hierarchical design is a prominent theme in many natural systems where mechanical efficiency is of importance. Here we establish the properties of a particular hierarchical structure, showing that high mechanical efficiency is found in certain loading regimes. We show that in the limit of gentle loading, the optimal hierarchical order increases without bound. We show that the scaling of material required for stability against loading to be withstood can be altered in a systematic, beneficial manner through manipulation of the number of structural length scales optimised upon. We establish the relationship between the Hausdorff dimension of the optimal structure and loading for which the structure is optimised. Practicalities of fabrication are discussed and examples of hierarchical frames of the same geometry constructed from solid beams are shown. 
\end{abstract}
\maketitle
\section{Introduction}
Hierarchical designs are found throughout nature where highly mechanically efficient load bearing structures are required \cite{Lakes}. Trabecular bone serves as a prime example where requirements for stiffness and strength are met through utilising structural hierarchy \cite{Huiskes}. Allometric scaling has been observed in the lattice-like sub-structure of the trabecular bone: in mammals, lattice connectivity increases and trabeculae thickness decreases with decreasing mass of animal \cite{Trab_Allo}. The structure of fossilised ammonites have long been appreciated to exhibit a fractal structure in their suture lines \cite{Saunders_2}. Although other driving factors have been proposed \cite{Per_Palm}, it has been persistently hypothesised that higher degrees of structural hierarchy are responsible for increased resilience to pressure bearing (for a given mass of construction material) \cite{Saunders_2, Oloriz, Long}. It has recently been shown that hierarchical and fractal-like suture joints can be used to tailor mechanical properties, load resistance and flaw tolerance \cite{Li}. Further examples include spider capture silk \cite{Du_spider, Zhang}, nacre \cite{Nac_1} and gecko setae \cite{Gao, Yao_1}, all exhibit hierarchical structures with geometric parameters tailored for different loading conditions. 

Recent theoretical works have found that efficient structures can be generated through a self similar design principle \cite{Farr_and_Mao, Ray_hol, Ray_pressure}. Under external pressure and gentle compressive loading the same tendencies are found: with decreasing load, the optimal number of hierarchical orders is found to increase and a tendency towards more slender components is observed \cite{Farr_and_Mao, Ray_hol, Ray_pressure, Ray_fab}. 

Advances in construction techniques have made it possible to fabricate designs with structural order on a wide range of length-scales. The construction of frames with a photosensitive polymer can be used in conjunction with other techniques such as electroless nickel plating and etching, to create frames with the same geometry but constructed from hollow, metallic tubing \cite{Schaedler, Jacobsen, Saleh}. Using such techniques, hierarchical metallic lattices, for example, have been created with structural order ranging from the nano- to the centimetre scale \cite{Schaedler}. Such techniques make it possible to design and create materials where beneficial properties of the macrostructure are bought about through the prudent choice of design parameters at structural length scales orders of magnitude smaller. 

Here, a hierarchical spaceframe design constructed from hollow tubes is analysed in full and its benefits over a solid beam construction are discussed. A spaceframe constructed from solid beams is created through rapid prototyping techniques showing the fabrication of the thin walled structure to be a plausible goal. The optimal number of hierarchical levels for a given loading is found and fractal dimension of the optimal structure is calculated. We discuss particular issues of practical importance when designing and fabricating hierarchical structures. 
\section{Theory}
\subsection{Solid beam and generation-0}\label{g-0}
To serve as a reference, we first consider the problem of obtaining the amount of material that is required, $V_{\text{req}}$, to construct a beam of length $L$, freely hinged at its ends, stable under a compressive load $F$. If we take an initially straight, solid, slender beam with circular cross section, constructed from an isotropic material, we see that the Euler buckling mode of the strut gives the first limit on stability. The load at which this instability is reached is given by:
\begin{equation}
F < \frac{\pi^2 Y I}{L^2}, \label{Euler_1}
\end{equation}
where $I$ is the second moment of area ($I = \pi r^4/4$ for a solid beam with circular cross-section) and $Y$ is the Young's Modulus of the material. 
Suitable non-dimensional variables for this problem are defined as,
\begin{eqnarray}
f \equiv \frac{F}{YL^2}, \label{fL} \\
v \equiv \frac{V_\text{req}}{L^3},\label{vL}
\end{eqnarray}
measuring loading and volume respectively. For a given $f$, setting $r$ such that the beam is on the point of instability due to Euler buckling gives us the minimum volume required for stability. This value of $r$ is found to be, 
\begin{equation}
r = L\left(\frac{4f}{\pi^3}\right)^\frac{1}{4}.
\end{equation}
Thus the minimum non-dimensional volume of material, $v$, required for stability for a given loading, $f$, can be expressed as:
\begin{equation}
v = 2\pi^{-\frac{1}{2}}f^\frac{1}{2}.
\end{equation}
As a comparison, it is noted that the volume of material required for stability under tension varies linearly with the loading parameter, or $v \propto f$. In all practical applications the non-dimensional
parameters $f$ and $v$ are much smaller than 1. Thus, to support a given magnitude of loading over a given distance requires less material if the support is under tension rather than compression. Furthermore, it is seen that splitting a given load over two tension members, each supporting half the load, has little consequence on the total volume of material required; on the other hand, as a result of the above scaling, the amount of material required increases greatly if multiple compression members are used to support a given load \cite{Cox}. 

If instead the circular beam is taken to be hollow, with thin walls, two restrictions are seen to apply to the loading. The first is given by Eq.~(\ref{Euler_1}) with $I = \pi[(r+t)^4-r^4]/4$ where $t$ is the thickness of the cylinder wall. Secondly, a short wavelength failure mode must be considered, Koiter buckling \cite{Koiter}, giving a second inequality:
\begin{equation}
F < \frac{2\pi Y t^2}{\sqrt{3(1-\nu^2)}},\label{Koiter}
\end{equation}
where $\nu$ is the Poisson ratio. Setting the geometry of the beam to be such that Euler buckling and Koiter buckling occur at the same value of loading, it is straightforward to show that
\begin{equation}
v = 2\left[\frac{3\left(1-\nu^2\right)}{4\pi^2}\right]^\frac{1}{6} f ^\frac{2}{3}.
\end{equation}
In the regime $f\ll1$ this change in scaling law represents a saving in material over the solid beam. In this work, the hollow cylinder will be referred to as the generation-0 structure. 
\subsection{Hollow generation-1 structure}
\begin{figure}
\begin{center}
\includegraphics[width = 3.30in]{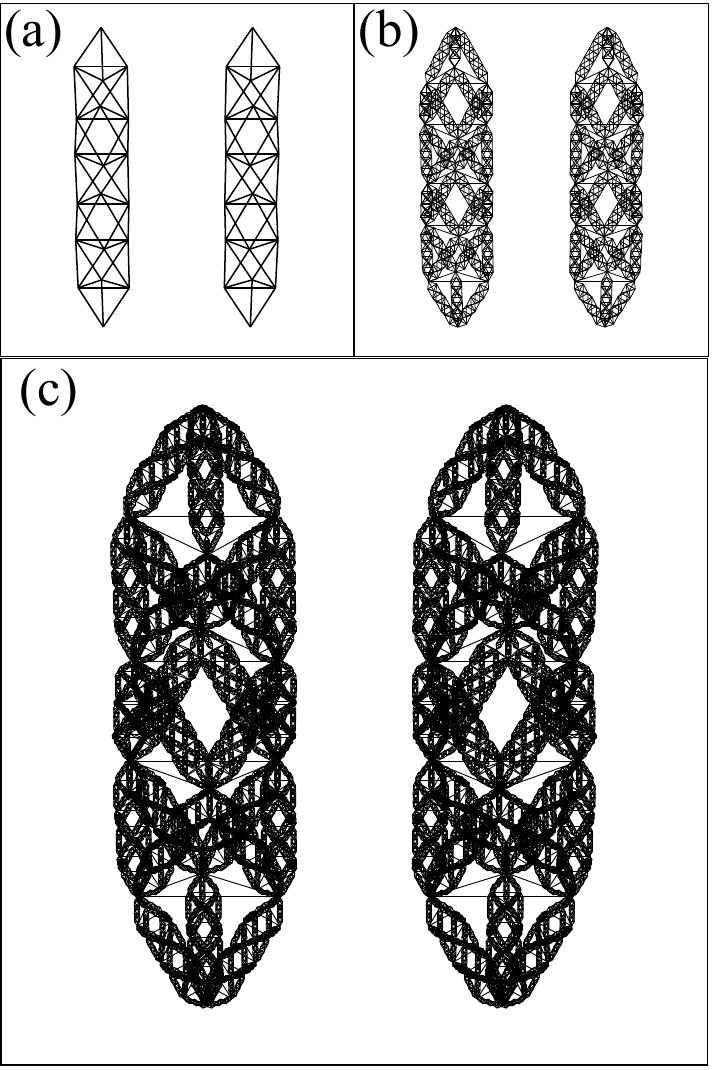}
\end{center}
\caption{Showing the progression to higher generations of the hierarchical structure. (a) depicts the simple spaceframe, (b) shows the space frame with 2 levels of hierarchy, while (c) shows a spaceframe with 3 levels of hierarchy. Images shown are stereographic: to see the 3-d image, hold the page 20-40cm away and stare ``through'' the paper until the images merge.\label{frame123}}
\end{figure}
The generation-1 structure is a simple spaceframe made up of $n$ octahedra which separate two end tetrahedra: the geometry of the spaceframe is shown in figure \ref{frame123} (a) with $n = 5$. Here we consider the component cylinders to be hollow with thin walls. If the length of the whole structure is defined as $L$, and the length of an individual component beam is $L_0$, then,
\begin{equation}
L = \sqrt{\frac{2}{3}}(n+2)L_0.\label{L}
\end{equation}
Assuming all beams in the structure to be made up of identical beams that exhibit
Hookean behavior for loading less than the Euler limit and whose spring constant is given by,
\begin{equation}
k_{0} = \frac{YA}{L},\label{k0}
\end{equation}
where $A$ is the cross-sectional area of the beam. For large enough $n$, the whole frame can be shown to have a bending stiffness, $YI$, and spring constant, $K$ given by:
\begin{eqnarray}
YI = BL^3_{0}k_{0},\label{stiffness}\\
K = \frac{36 k_{0}}{11n+43},
\end{eqnarray}
where $B$ is a constant found to be $B = 0.245\pm0.001$ \cite{Farr_and_Mao}. 
If the structure is oriented such that the end points of the tetrahedra are aligned along the $z$-axis in Cartesian coordinates, then on loading these end points with a force $F$ in a compressive manner, it is found that that all beams parallel with the $x-y$ plane are under tension. Assuming $n\ge2$ the beams under tension making up the end tetrahedra support a load of 
$\frac{F}{2\sqrt{6}}$ while other tension members support a load of $\frac{F}{3\sqrt{6}}$. 
It is found that all other beams support a compressive load. The beams connected to the end points are acted on by a force of
\begin{equation}
F_{0} = \frac{F}{\sqrt{6}},\label{F0}
\end{equation}
while all other beams under compression take half this load. In the generation-1 frame, there are 3 failure modes: Koiter buckling of the individual beams and Euler buckling of both the composite frame and the individual beams. The three parameters that we wish to optimise over are $r$, $t$ and $n$. We proceed by defining 
\begin{equation}
f_0 \equiv \frac{F_{0}}{YL_{0}^2},\label{f0}
\end{equation}
and stating that the beams connected to the loading points of the structure are on the point of simultaneous failure due to both Euler and Koiter buckling. Through use of Eqs.~(\ref{Euler_1}, \ref{Koiter}, \ref{F0} \& \ref{f0}) it follows that:
\begin{eqnarray}
t &=& L_{0}\left[\frac{\sqrt{3\left(1-\nu^2\right)} f_0}{2\pi}\right]^\frac{1}{2},\label{t}\\
r &=& L_{0}\left[\frac{2f_0}{\pi^5\sqrt{3(1-\nu^2)}}\right]^\frac{1}{6}.\label{r}
\end{eqnarray}
Then, using Eqs.~(\ref{Euler_1}, \ref{L} - \ref{stiffness} \& \ref{F0} - \ref{r}) and setting the whole spaceframe to be on the point of Euler buckling, it is found that 
\begin{equation}
n = -2 + \Bigg\lfloor\frac{6^\frac{1}{4}\pi^\frac{5}{6}B^\frac{1}{2}\left[3\left(1-\nu^2\right)\right]^\frac{1}{12}f_0^{-\frac{1}{6}}}{2^\frac{2}{3}}\Bigg\rfloor,
\end{equation}
where $\lfloor\cdot\rfloor$ is the floor function. Then, using Eqs.~(\ref{L} \& \ref{F0}) it is found that,
\begin{equation}
f = \frac{3\sqrt{6}}{2}(n+2)^{-2}f_0.\label{f_G1}
\end{equation}
Using Eqs.~(\ref{L}, \ref{t} \& \ref{r}), the non-dimensional volume is found to be,
\begin{equation}
v = 27\sqrt{6}\frac{(n+1) f_0^\frac{2}{3}\left[3\left(1-\nu\right)\right]^\frac{1}{6}}{\pi^\frac{1}{3}2^\frac{4}{3}\left(n+2\right)^3}, 
\end{equation}
thus, through use of Eq.~(\ref{f_G1}), 
\begin{equation}
v \propto f ^\frac{3}{4} + O(f^\frac{7}{8}).
\end{equation}
This expression represents a gain in efficiency over both the solid and hollow beams in the limit $f\ll1$. For comparison a spaceframe constructed from solid beams scales as $v\sim f^\frac{2}{3}$. Thus it is seen, in the limit of gentle loading, the structure presented here is more efficient. 
\subsection{Generation-$G$ optimisation}
\begin{figure}
\begin{center}
\includegraphics[width = 3.3in]{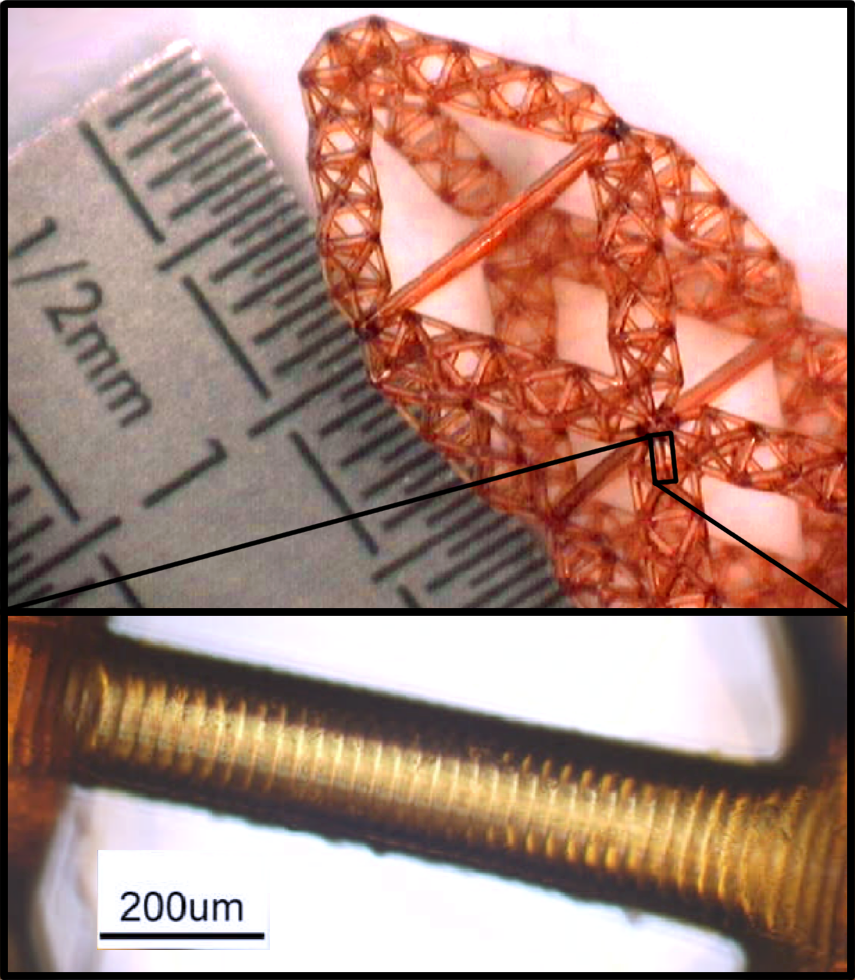}
\end{center}
\caption{\textcolor{red}{(Color online)} Showing the upper tetrahedron and first octahedron of a generation-2 hierarchical structure constructed through a rapid prototyping technique. This structure was created through use of EnvisionTEC Perfactory machine. Inset shows the layering effect of the rapid prototyping procedure. The layer thickness of the structure shown is approximately 25$\mu$m. The material used in the construction of this structure is EnvisionTec R05 \cite{envisionTEC}.\label{structure}}
\end{figure}
The generation-$G$ structure can be created through an iterative procedure. In creating the generation-1 structure, the simple, hollow beam that makes up the generation-0 structure is replaced with a spaceframe. It is an analogous step that takes us from the generation-1 structure to the generation-2 structure: all simple beams in the structure that are loaded under compression are replaced by (scaled) generation-1 frames. Thus, it is noted, a generation-$G$ constructed from hollow tubes has $G+2$ characteristic length scales upon which it could fail. The notation used here will follow that in Ref.~\cite{Farr_and_Mao}:] a given property of the structure that is recurrent on different hierarchical levels of the structure will be denoted $X_{G,i}$, which represents the property $X$ on the $i$-th level in a generation-$G$ structure ($i=0$ and $i=G$ denotes the smallest and largest length scale in the structure respectively).
The generation-1, 2 and 3 structures are shown in stereographic form in figure \ref{frame123}. Shown in figure \ref{structure} is the upper tetrahedron and octahedron of a generation-2 spaceframe constructed through rapid prototyping techniques. 

The properties of any (sub)frame can be related to the smallest component beams through expressions involving $\{n_{G,i}\}$. These expressions are dependent only on the geometry of the spaceframe and are given by:
\begin{eqnarray}
F_{G,i} &=& 6^\frac{i}{2}F_{G,0}\label{Fi}\\
L_{G,i} &=&\left( \frac{2}{3} \right)^\frac{i}{2} \prod_{j=1}^i \left(n_{G,j}+2\right)L_{G,0},\label{Li}\\
k_{G,i} &=& 36^i \prod_{j=1}^i \left(11n_{G,i}+43\right)^{-1} k_{G,0},\label{ki}\\
YI_{G,i} &=& B\left(\frac{2}{3}\right)^\frac{3\left(i-1\right)}{2}\prod_{j=1}^{i-1} \frac{\left(n_{G,j}+2\right)^3}{11n_{G,j}+43} L_{G,0}^3k_{G,0}\label{YIi}
\end{eqnarray}
where $k_{G,i}$ is the effective spring constant of all (sub)structures of length $L_{G,i}$, and $F_{G,i}$ is the applied compressive load to each substructure of length $L_{G,i}$.

It is seen that to avoid Euler buckling at each hierarchical 
length scale, the constraint
\begin{equation}
F_{G,i} < \frac{\pi^2YI_{G,i}}{L_{G,i}^2}\label{F_Gi}
\end{equation}
must be imposed for all $i$. Given that the smallest beams are made of hollow tubes, the possibility of Koiter buckling must be taken into account. This constraint on 
loading provides us with the inequality
\begin{equation}
F_{G,0} < \frac{2\pi Y t^2}{\sqrt{3\left(1-\nu^2\right)}}.
\end{equation}
\begin{table}[htdp]
\caption{Showing the optimal parameters for both a hollow and solid construction hierarchical frame. The loading for which this frame is optimal is $F$ = 1kN, $\nu = 0.29$, $Y = 210$GPa}
\begin{center}
\begin{tabular}{c||c|c|c|c}
Generation & $n_{G,G}$ &$n_{G,G-1}$ &$n_{G,G-2}$ & Mass (kg)\\
\hline
\hline
\bf{Hollow - 0} &-&-&-&\bf{1421}\\
Solid - 0&-&-&-&$7.9\times10^4$\\
\hline
\bf{Hollow - 1}&\bf{44}&-&-&\bf{487}\\
Solid - 1&140&-&-&2920\\
\hline
\bf{Hollow - 2}&\bf{22}&\bf{22}&-&\bf{439}\\
Solid - 2&46&47&-&1790\\
\hline
\bf{Hollow - 3}&\bf{13}&\bf{14}&\bf{14}&\bf{533}\\Solid - 3&23&23&24&2180\\
\end{tabular}
\end{center}
\label{default}
\end{table}%
The parameters over which we optimise are $r$, $t$ (which are assumed to be constant over the generation-1 structure), and $\{n_{G,i}\}$. Defining the geometry such that Euler buckling and the short wavelength Koiter buckling occur simultaneously in the beams of length $L_{G,0}$, through use of Eqs.~(\ref{Koiter}, \ref{f0} \& \ref{F_Gi}) with $i=0$, it can be shown that $r$ and $t$ are given by:
\begin{eqnarray}
t &=& L_{G,0}\left[\frac{\sqrt{3\left(1-\nu^2\right)} f_0}{2\pi}\right]^\frac{1}{2},\label{tG}\\
r &=& L_{G,0}\left[\frac{2f_0}{\pi^5\sqrt{3(1-\nu^2)}}\right]^\frac{1}{6}.\label{rG}
\end{eqnarray}

Using these experssions and Eq.~(\ref{k0}) it can be shown that
\begin{equation}
k_{G,0} = L_{G,0}Y \left[\frac{4f_0^2\sqrt{3\left(1-\nu^2\right)}}{\pi}\right]^\frac{1}{3}.\label{kG0}
\end{equation}
Then using Eqs.~(\ref{f0} - \ref{r}, \ref{Li}, \ref{YIi}, \ref{Fi} \& \ref{F_Gi}), setting all (sub)frames to be on the point of failure due to Euler 
buckling, it can be observed that, 
\begin{equation}
n_{G,1} = -2 + \Bigg\lfloor\frac{6^\frac{1}{4}\pi^\frac{5}{6}B^\frac{1}{2}\left[3\left(1-\nu^2\right)\right]^\frac{1}{12}f_0^{-\frac{1}{6}}}{2^\frac{2}{3}}\Bigg\rfloor,\label{n1}
\end{equation}
and, for $i>1$,
\begin{eqnarray}
n_{G,i} = - 2 + \Bigg\lfloor\left\{ \frac{\sqrt{6}}{2^\frac{4}{3}}\pi^\frac{5}{3}B\left[3\left(1-\nu^2\right)\right]^\frac{1}{6}f_0^{-\frac{1}{3}} \nonumber\right. \\ \left.12^{i-1}\prod^{i-1}_{j=1} \frac{n_{G,j}+2}{11n_{G,j}+43}\right\}^\frac{1}{2}\Bigg\rfloor.\label{ni}
\end{eqnarray}
In this calculation, the spring constant of the simple beams under tension at any given hierarchical level are chosen to be equal, they are set as that of the spaceframe of the same length. To achieve this, using Eqs.~(\ref{Li}, \ref{ki}, \ref{tG}, \ref{rG} \& \ref{kG0}) we see that the radii, $s_{G,i}$, of the tension resisting beams of length $L_{G,i}$ are, 
\begin{equation}
s_{G,i} = (12\sqrt{6})^i\prod^i_{j=1}\frac{n_{G,j}+2}{11n_{G,j}+43}r,\label{sGi}
\end{equation}
and $t$ remains constant for the whole structure. For $G>1$, it is found that,
\begin{eqnarray}
f = \left(\frac{27}{2}\right)^\frac{G}{2}f_0\prod^G_{j=1}\left(n_{G,j}+2\right)^{-2} \label{f_opt}\\
v = \left(\frac{9\sqrt{6}}{2}\right)^G\frac{f_0^\frac{2}{3} \left[3\left(1-\nu^2\right)\right]^\frac{1}{6} }{2^\frac{1}{3}\pi^\frac{1}{3}}\prod^{G}_{k=1}\frac{n_{G,k}+1}{(n_{G,k}+2)^3} \nonumber \\ 
\left[3+ \sum^{G-1}_{q=1}4^q\prod_{j=1}^{q} \frac{(n_{G,j}+2)^2}{(11n_{G,j}+43)(n_{G,j}+1)}\right].\label{v_opt}
\end{eqnarray}
\begin{figure}
\begin{center}
\includegraphics[width = 3.2in]{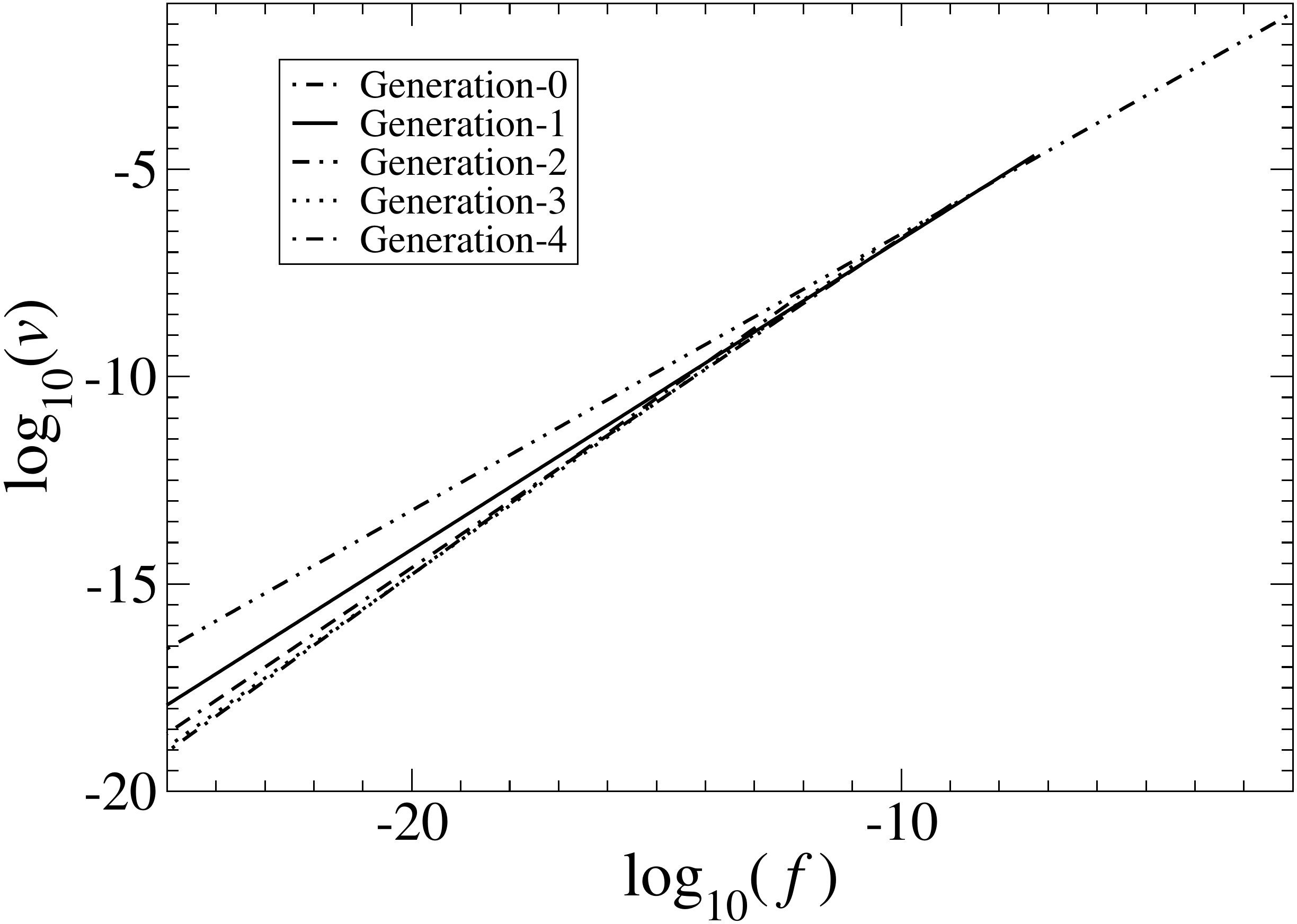}
\end{center}
\caption{Volume required for structural stability against loading for which the 
structure is optimised, showing generation-0 to 4. Higher generations become optimal as the loading parameter, $f$, decreases.\label{fvG}}
\end{figure}
\begin{figure}
\begin{center}
\includegraphics[width = 3.3in]{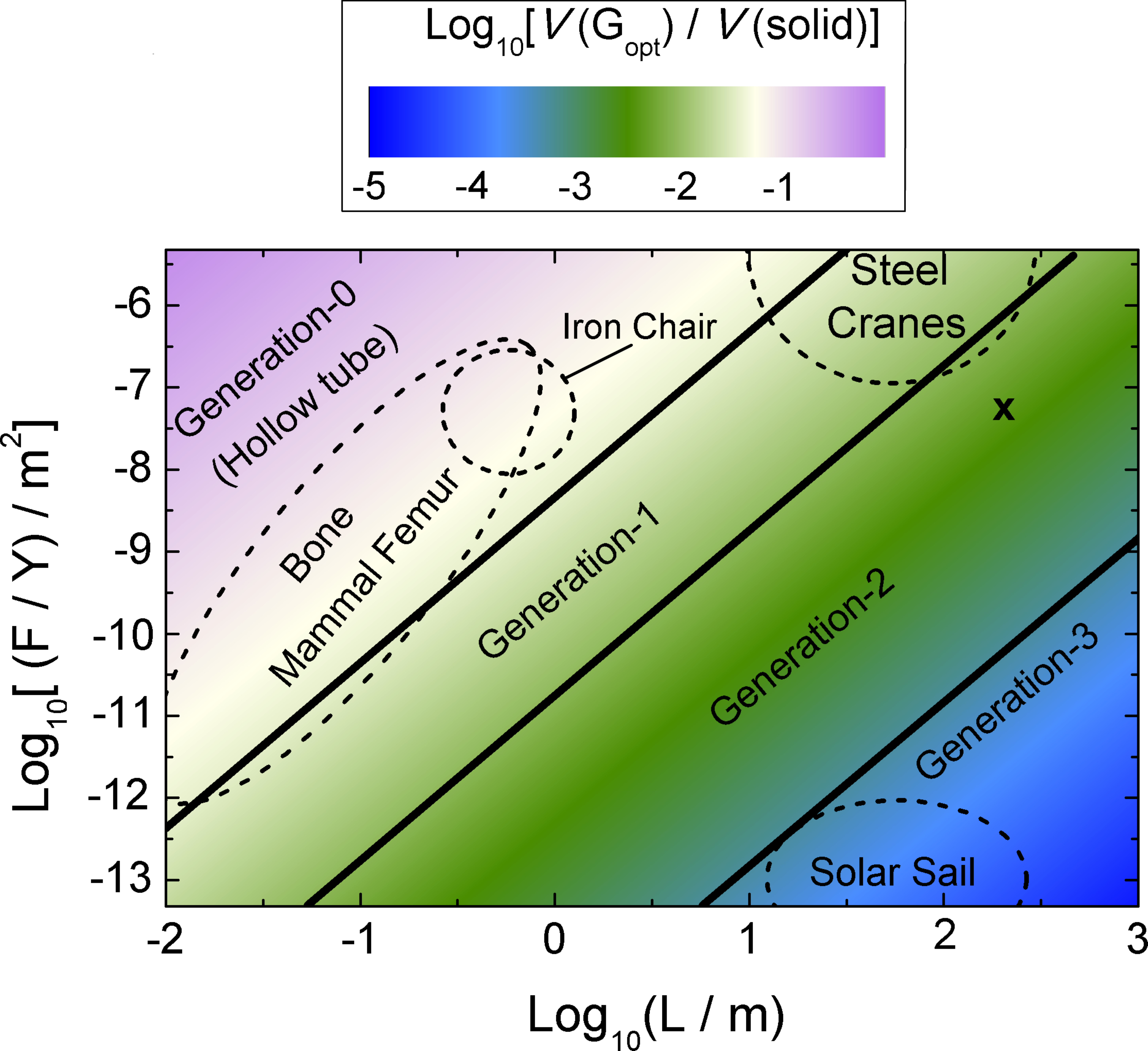}
\end{center}
\caption{\textcolor{red}{(Color online)} The material saving through use of a generation-$G$ hollow tube structure when compared to a solid beam. The plot is valid for a material with Poisson ratio of $\nu \approx 0.3$. The progression of optimality for higher generation designs is clearly shown with the increase in length, $L$, or decrease in force, $F$, for a given Young's Modulus, $Y$. Also depicted are regions showing typical parameters for some compression bearing structures: approximate regions for steel crane booms \cite{liebherr}, iron chair legs, solar sail compression beams \cite{Solar_Sail} (from an arbitrary stiff material, $Y>100$GPa) and mammal femurs withstanding only static loads \cite{Trab_Allo}. Also shown is the positioning of the test problem investigated in the text and in table \ref{default} ($F = 10$kN, $L = 200$m and $Y = 210$GPa). \label{complexity}}
\end{figure}
\begin{figure}
\begin{center}
\includegraphics[width = 3.2in]{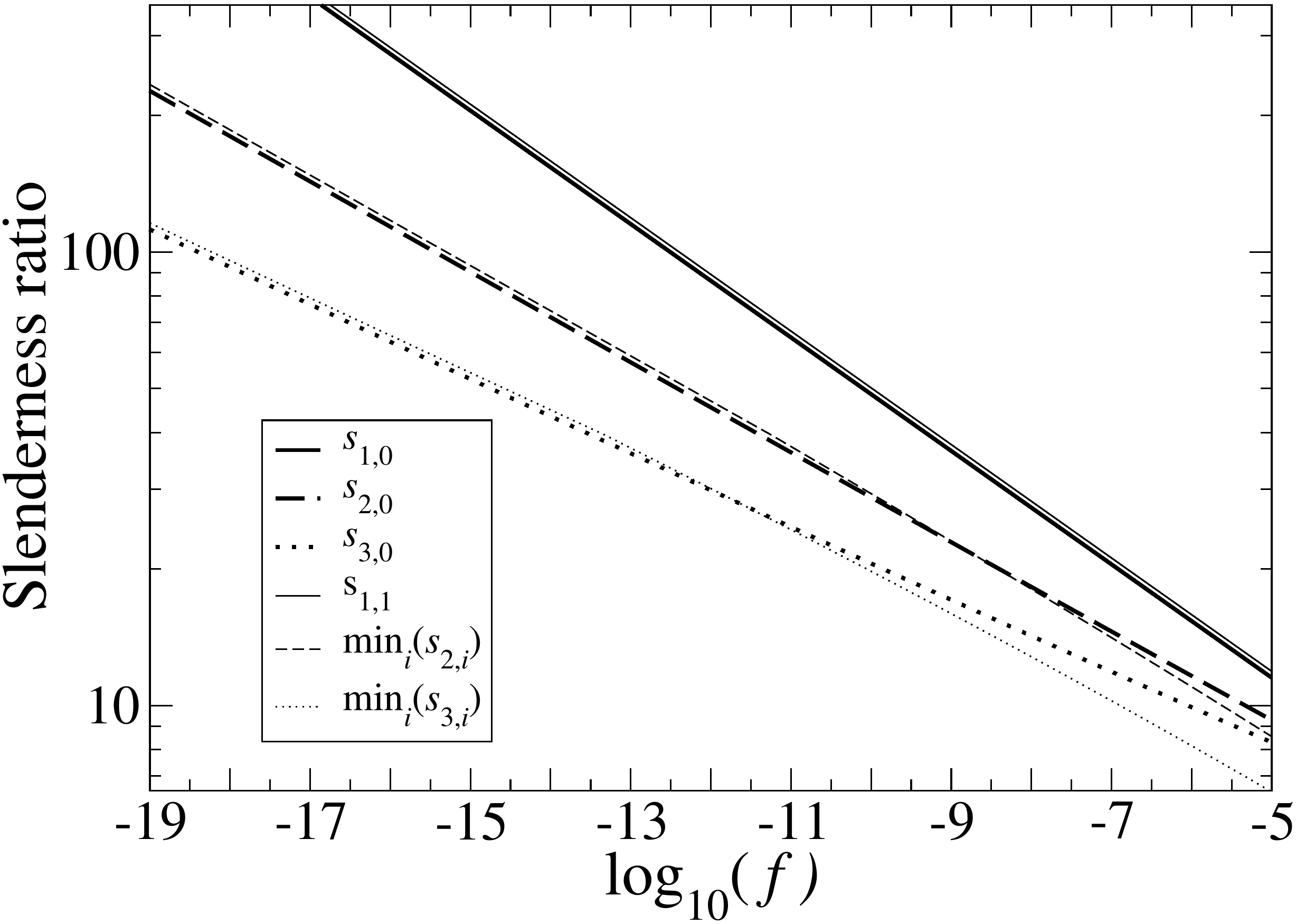}
\end{center}\caption{Variation of both the aspect ratio for the smallest beams and the minimum aspect ratio of all (sub-)frames for generation-1 to 3 with respect to loading parameter. Showing increasing aspect ratio as the loading for which the structure is optimised is decreased.\label{asp}}
\end{figure}
To obtain the former equation, Eqs.~(\ref{fL}, \ref{f0}, \ref{Fi} \& \ref{Li}) were 
used, and in the latter, Eqs.~(\ref{vL},  \ref{Li}, \ref{tG} \& \ref{rG}).
The scaling of material required to make a stable structure 
out of hollow tubes, to leading order, is therefore shown to obey:
\begin{equation}
v \approx \kappa_\text{hol}(G) f^\frac{G+2}{G+3}.\label{vf_scaling}
\end{equation}
Combining Eqs.~(\ref{f_opt} \& \ref{v_opt}) and eliminating $f_0$ a full expression for the volume required for stability under a given load can be obtained, and this is plotted in figure \ref{fvG}, where the scaling of Eq.~(\ref{vf_scaling}) is seen to dominate. For $f \ll 0$, this design shows that considerable gains in efficiency are possible through increasing the hierarchical order of the structure. In the limit $f \rightarrow 0$ the scaling of material required for stability against loading to be withstood is seen to tend to that found for a simple beam under tension. For a given material, $\kappa_{2}(G)$ increases with increasing $G$. Thus, for all non-zero values of loading, the optimal generation is found to be finite. The progression of the optimal generation of the hierarchical frame is shown in figure \ref{complexity} where the material saving and optimal generation are plotted for various values of $F/Y$ and $L$. We see that, for a given $Y$, low $F$ and high $L$ (or small $f$) lead to higher generation numbers being more efficient. 

The optimisation procedure described above results in a structure that sets
\begin{eqnarray}
L_{G,0} &\propto& \sqrt{r L_{G,1}} ,\\
r &\propto& \sqrt{tL_{G,0}},
\end{eqnarray}
and in the limit $f\rightarrow0$, $L_{G,i}$ is approximated by
\begin{equation}
L_{G,i} \propto \sqrt{L_{G,i+1} L_{G,i-1}}\quad \text{for} \quad 2\le i \le G-1 .
\end{equation}
The analysis above assumes that on all length scales elastic failure is the active failure mode. Plotted in figure \ref{asp} is the slenderness ratio for the smallest beams and the minimum aspect ratio of all the spaceframes in the structure, as defined by the expressions:
\begin{eqnarray}
s_{G,0} &\equiv& \frac{L_{G,0}}{r}, \\
\min_{i>0} \left(s_{G,i}\right) &\equiv& \min_{i>0}\frac{d_{G,i} }{L_{G,i}},\label{di}
\end{eqnarray}
where $d_{G,i}$ is the maximum distance of any material making up a spaceframe of length $L_{G,i}$ from its neutral axis. As the loading parameter becomes smaller, the slenderness ratio increases, thus it is likely in the regime where maximal gains from the hierarchical construction are found, the elastic (as opposed to plastic) failure mode is dominant. 
\section{Fractal Dimension}
\begin{figure}
\begin{center}
\includegraphics[width= 3.2in]{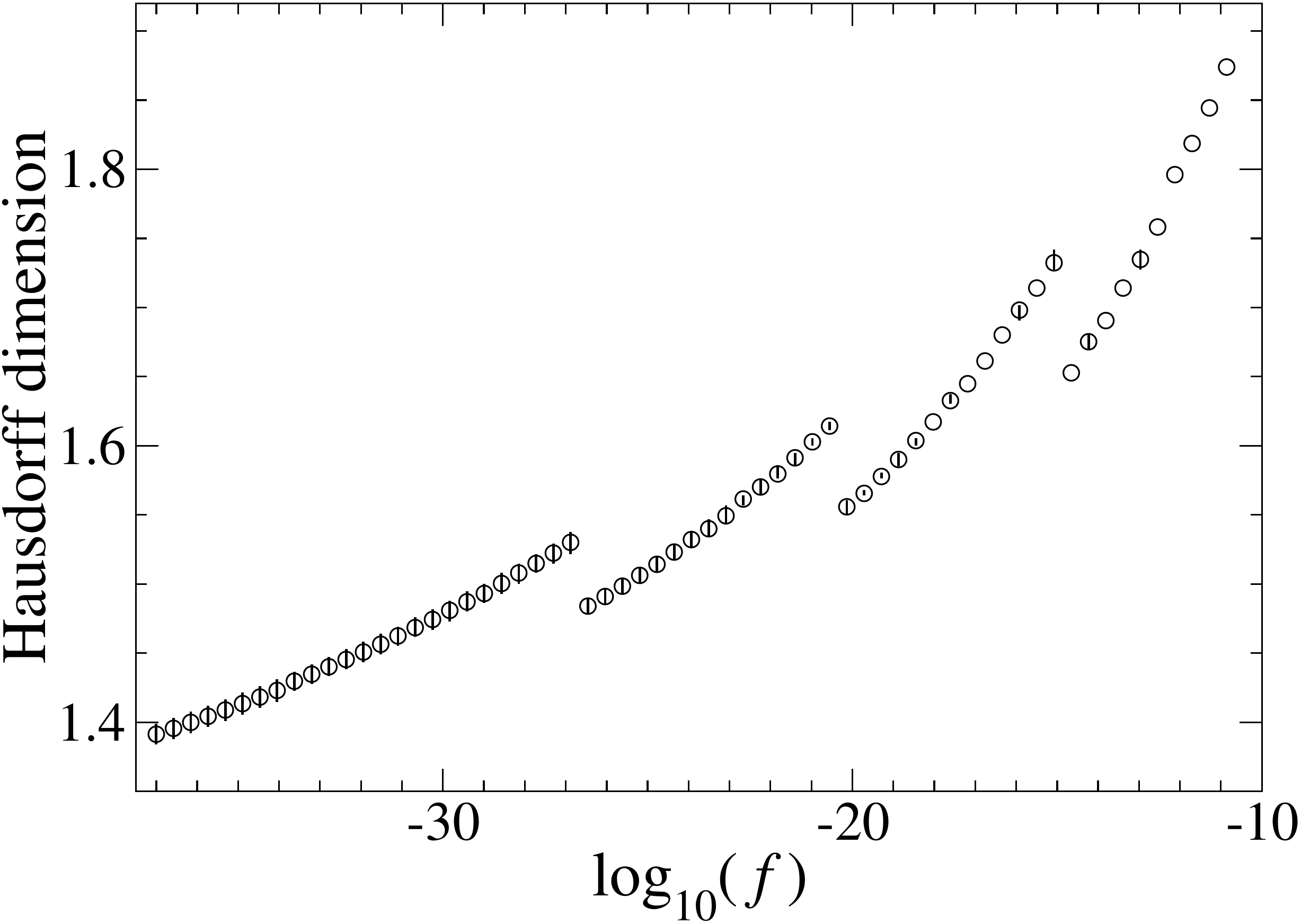}
\end{center}\caption{The fractal dimension for the optimal structure plotted against the loading for which the structure is optimised. Bar shows the variation of the Hausdorff dimension over all appropriate length scales while the circle shows the average Hausdorff dimension of the structure. Discontinuities in dimension represent transitions of optimality from one generation to another. \label{frac_dim}}
\end{figure}
The structures described above are hierarchical over a certain range of length scales. Within this range, one can calculate an effective Hausdorff dimension, $D$, through considering the self similarity of the structure at different hierarchical levels. It is found that it is dependent on $n_{G,i}$ and is given by the following expression:
\begin{equation}
 D = \frac{2\log_{10}\left[6\left(n_{G,i}+1\right)\right]}{\log_{10}\left(\frac{2}{3}\right)+2\log_{10}\left(n_{G,i}+2\right)}.\label{frac_dim_eq}
\end{equation}
The values shown in figure \ref{frac_dim} are confirmed through a box-counting technique. The box counting method is used with a set of cubes with side length of $2r$ or $d_{G,i} \text{ where } \: i \in \:[1,G]$ ($d_{G,i}$ takes the same values as in Eq.~(\ref{di})). For small enough box sizes (below the range of length scales where the structure is hierarchical), a structure optimised for a finite force will have a Hausdorff dimension of 3. For suitably small but finite values of loading however, the set $\{d_{G,i}\}$ will yield a non-trivial Hausdorff dimension. As a result of the variation of $n_{G,i}$ with $i$, the fractal dimension of the optimal structure described above is not a constant over all length scales. The upper and lower bounds for the fractal dimension can, however, be found. These bounds are shown in figure \ref{frac_dim} where they are plotted against the loading parameter for which the structure is optimised. A sub-optimal structure with a constant Hausdorff dimension could be created by setting $n_{G,i} = n_{G,G} \;\forall\;i$ where $n_{G,G}$ is taken from the optimised structure. Such a structure, with $t$ and $r$ taken from Eqs.~(\ref{tG}) and (\ref{rG}) respectively, would be stable for loading greater than $f$ and would attain the upper bound in Hausdorff dimension shown in figure \ref{frac_dim}. In the limit of $f\rightarrow 0 $ it is seen from Eqs.~(\ref{n1}, \ref{ni}, \ref{frac_dim_eq}) that the limit of the fractal dimension tends to 1. 

\section{Fabrication of hierarchical structures}

With the development of novel fabrication techniques the engineering challenge in creating these structures is not insurmountable. In most terrestrial applications, it is found the optimal level of hierarchy, for this structure, will not exceed 3. This puts some restrictions on, but does not negate, the engineering challenge. 

We have fabricated fractal compression members using a modified EnvisionTEC Perfactory\circledR~type III mini system from a photosensitive polymer, EnvisionTEC R05 \cite{Ray_fab}. This mask-projection based photopolymerisation system has a $2800\times2100$ pixel digital light processing projector allowing a resolution of 5$\mu\rm{m}$. The structure shown in figure 2 was first modelled in 3D as an STL file, before being split into numerous thin layers and stored as a job file using Perfactory RP proprietary software. These layers are visible in the final manufactured structure, see figure \ref{structure}. Light with wavelength approximately 475nm is then passed through the projector and focused onto the resin surface for polymerisation of the exposed areas. The sample is then washed using ispropanol in an ultrasonic bath and left to dry. A postcuring procedure is followed using an EnvisionTEC Otoflash System to harden the material. An alternative material, EnvisionTEC RC25 (Nanocure), has also been used to create frames of the same geometry using the same fabrication procedure but without the necessity for postcuring \cite{Ray_fab}.

The structure shown in figure \ref{structure} is a generation-2 hierarchical frame with $n_{2,1} = 5$ and $n_{2,2} = 4$. The smallest beams in the structure have radii of approximately 0.15mm and lengths of 1.35mm. The layer thickness of the growth was 25$\mu$m. 

Mechanical testing of the structure presented here has been undertaken \cite{Ray_fab} and good agreement between the structure's performance and finite element simulations is found. It is noted that, in this case, the bending moments induced at the beam ends due to deformation result in failure at lower loads than predicted in the theory presented here. It is also noted that the ``slender beam'' approximation used here is not well met by structures fabricated to date. 

\section{Conclusions}
We have shown that through a hierarchical design principle a highly efficient compression bearing structure can be created. Analysing all possible modes of failure, at each length scale, we have shown that the scaling of volume of material required for stability against a given loading can be systematically varied in an advantageous manner. We have shown that the use of hollow, rather than solid beams, changes the scaling in a manner analogous to increasing the generation number by 1. More generally it is noted that for hierarchical structures optimised for gentle compressive loading here and in Refs.~\cite{Farr_and_Mao, Farr_Efficient_Shell}, a structure with $n$ characteristic length scales of failure obeys a relationship of $v\propto f^\frac{n}{n+1}$. We have also shown the dependence of fractal dimension of the optimal structure on the applied load at failure. The dependence on loading of the optimal number of levels of hierarchy for this structure has also been obtained. 

Further optimisation of the structure is possible: at every hierarchical level, there exist two different loading conditions for beams/sub-frames under compression. Despite this, in the work presented here, all beams and sub-frames at a particular hierarhical level are equivalent; variation of sub-frame charateristics, optimising each one for its particular loading would result in a more efficient structure. 

The use of these hierarchical structures will be dependent on cost of production and robustness of the structure in their intended use. The potential trade-off between mechanical efficiency and robustness of hierarchical structures must be investigated further. It is noted that the structure presented above is minimally rigid. Thus, for this particular structure, modeled with freely hinged joints, removal of a single beam (or subframe) will result in collapse of the structure at all larger lengthscales. In the case of non-freely hinged joints, some rigidity will be maintained however a transition from stretching to bending dominated regimes will occur. 


Finally it is noted that the smallest possible building blocks for these hierarchical designs are single and multi wall carbon nanotubes. It has been shown that both Koiter and Euler buckling of these tubes are closely approximated by Eqs.~(\ref{Euler_1}) and (\ref{Koiter}), up to a prefactor in the case of multi-walled carbon nanotubes \cite{CNTs}. Thus it is expected that the analysis shown previously will still hold. Alternative structural elements include hollow nanotubes constructed through atomic layer deposition \cite{Ikkala_tubes}. Ultimately, molecular self-assembly may offers a fabrication method for these intricate hierarchical materials with structural features from the nanoscale up \cite{self_assembly, inorganic_rods}. 
\section{Acknowledgements}
The authors would like to acknowledge the work of Joel Segal, Ranbir Singh and Fred Grillet, from the Department of Mechanical, Materials and Manufacturing Engineering, University of Nottingham also that of Mark Strickland of the Faculty of Engineering at the University of Nottingham. 

\appendix

\section{Stereolithography files}
The fabrication of these intricate structures has become possible through advances in 3-d printing. With the increasing availability and resolution of commercial printers, the authors have made some example {\bf st}ereo{\bf l}ithography (.stl) files freely available online \cite{Yongs}.

\end{document}